         \newcommand{\lb}{\label}
         \newcommand{\nbc}{\nonumber}
         \newcommand{\nb}{\nonumber \\}
         \newcommand{\qb}{\begin{equation}}
         \newcommand{\qba}{\begin{eqnarray}}
         \newcommand{\qbas}{\begin{eqnarray*}}
         \newcommand{\qe}{\end{equation}}
         \newcommand{\qea}{\end{eqnarray}}
         \newcommand{\qeas}{\end{eqnarray*}}

         \newcommand{\rep}[1]{\protect\ref{#1}}

 \documentclass[12pt]{elsart}      
\usepackage{epsfig}
\begin{document}
\begin{frontmatter}
%
\title{chemical dynamics versus transport dynamics in a simple model} 

\author{H. Lustfeld}
\address{
 $^1$ Forum Modellierung and Institut f\"ur Festk\"orperforschung,\\
Forschungszentrum J\"ulich,
  D $52425$ J\"ulich, Germany}
and
\author{Z. Neufeld}
\address{Department for Atomic Physics, E\"otv\"os University, 
Puskin u. 5-7, H-1088
Budapest, Hungary}
\begin{abstract}
Reaction equations of homogeneously mixed pollutants 
in the atmosphere can lead to non-stationary periodic solutions.
It is important to know in which
respect these solutions are modified under the influence of the atmospheric
currents. We investigate this question in a very simple model: The
reaction equations are modeled by the equations of the brusselator and the
currents are represented by an isolated vortex. In the limit of high
vortex currents we find again the homogeneous solutions whereas for
smaller currents complicated spatial and temporal patterns emerge. The role of 
the diffusion as a singular perturbation is investigated.

\end{abstract}
\end{frontmatter}
\section{INTRODUCTION}
\label{intro}

The concentrations of pollutants in the atmosphere depend on the
atmospheric currents and the reaction equations between the pollutants.
(Moreover they depend on cloud formation, humidity, ice etc, influences
that will not be discussed here.)

When dealing with this problem the currents are usually replaced by their
average and mixing is modeled by introducing
'turbulent diffusion'\cite{Csa73}\cite{MoYa73}\cite{Pan85}\cite{NePaFi95}. 
The problem of such
approximations is that the chemical reactions depend on the {\em local}
concentrations and not on averaged ones. A further disadvantage of this
scheme is that the inserted turbulent diffusion is orders of 
magnitudes higher than the
molecular diffusion. 
This is questionable
because diffusion is a
singular\cite{Tur52}\cite{Bel64} and sensitive perturbation as will be shown below.

If the atmospheric currents
lead instantaneously to a homogeneous mixing of the pollutants, all 
concentrations are
obtained from the chemical reaction equations alone. 
The solutions of these reaction
equations 
need not approach a time-independent limit (fixed point for the dynamical
system) but
periodic fluctuations are possible as has
recently been shown in a model system containing six pollutants and two
pollutant sources\cite{PoLu96}\cite{KroPo97}\cite{Lu97}. 
A particular feature was the result that
pollutant concentrations may 
change by an order of magnitude within a few days. 

The solutions of the reaction equations become perturbed and then modified
as soon as the atmospheric currents do not lead to a complete mixing any
more. This situation could be
modeled by adding to the reaction equations the above mentioned 
averaged currents plus a turbulent
diffusion term that is
sufficiently high. But according to the remarks made earlier
a more appropriate approach is to assume that the
diffusion is weak
and to incorporate the transport equations in 
the reaction equations  directly. 

The temporal evolution of concentrations of $n$ reactants are
described by the reaction-advection-diffusion equations:
\qb 
\frac{\partial}{\partial t} c_i =
f_i(c_1..c_n,{\bf x},t)-{\bf v}({\bf x},t)\cdot 
\nabla c_i+\epsilon_i \Delta c_i,\;\;
i=1,..,n ,\lb{startEq}
\qe
where the functions $f_i$ describe the chemical reactions and 
pollutant sources. These functions 
depend explicitly on the spatial coordinate ${\bf x}$ if the 
reactant $i$
is produced by a localized source. The second term represents 
the advection of
pollutants by the velocity field 
${\bf v}({\bf x},t)$ and the last one is the diffusion
term. The coefficient $\epsilon$ is small and
in a first approximation we neglect the diffusion. Later on we
will investigate the addition of diffusion in our model
very carefully.

In general the equations eq.(\rep{startEq}) are rather involved.
In this paper we choose the  simplest possible system describing nontrivial
reactions between constituents subjected to a simple current which is
a solution of Euler's equation\cite{PeTeTo95}. 
The reaction equations are taken from
the brusselator\cite{Ha83}. Depending on the parameters the concentrations
converge to a fixed point or limit cycle in the homogeneous case.
Typically periodic solutions with sharp peaks occur. They are analogous
to those obtained in the model of
\cite{PoLu96} where
the
chemistry of tracer constituents in the troposphere were described.

For
maintaining the chemical reactions a 'pollutant source' is required.
We choose a 
point source pouring out
pollutants of sort $X$ placed in the velocity field
of a two-dimensional isolated point vortex. The pollutant $X$ decays, partly 
into
harmless substance $E$ partly into a second 
constituent $Y$ that 
autocatalytically reacts with $X$ again according to 
the well known reactions of the brusselator

\qba
X &\rightarrow & E\nb
X &\rightarrow & Y \lb{cRea}\\
2X +Y &\rightarrow & 3X\nbc
\qea
 
Since molecular diffusion does not lead to a spreading on a macroscopic
scale 
the vortex distributes the pollutants 
along the circular streamline containing the point source and thus
the model becomes one dimensional. We note that this 
advection-reaction problem can be seen as an
extension of the simple brusselator containing it as a limit
limiting
case
for high vortex strengths (or when the source is close to
the vortex center). 

Next we discuss the distribution of concentrations on the circle. 
When the time $T$, needed for one circulation is small, the 
solutions become very similar to the
reaction equations with a homogeneously distributed source. On the
other hand, if $T$ is larger than a threshold value
for an observer moving with the
fluid, the distribution becomes periodic with period $T$, whereas for an 
observer at rest the distribution is
stationary. When $T$ decreases
the distributions become rather different: 
Time dependent solutions
with period $nT$,
($n =2,3 ...$), 
quasiperiodic and chaotic
solutions are detected. All these lead to time dependent distributions that
are infinitely
degenerate on the circle and therefore depend on the initial
distribution. Even if that is smooth, steps in the distributions
occur, in the chaotic case on {\em each scale}. Therefore diffusion
is a singular perturbation and switching it on along the streamline leads to
drastic effects. For the period $2T$ case, the addition of diffusion
can be understood completely. We find two time scales. The first, 
which is of the
order of $1$\footnote{In this paper we use non-dimensional units.}
leads to coarse graining on the order of $\sqrt{\epsilon }$, 
the second which is of the
order of $e^{\beta /\sqrt{\epsilon }},\; \beta  =\mathcal{ O}(1)$ removes 
all the degeneracies of the solutions
leading to a distribution that is smooth apart from one
step. This step is intrinsic and would appear for a 
source of finite size along the streamline as well.

For parameters, that without diffusion lead
to periodic solutions of higher period we observe 
pattern formation and spatiotemporal chaos. 
All these 
solutions have nothing
in common with the case of homogeneous mixing we started with. 

In section II we present the model, in section III we discuss its properties
without diffusion, in section IV we concentrate on the role
of diffusion. The conclusion ends the
paper.

\section{THE MODEL}
\label{sec2}

The reaction processes, eq.(\rep{cRea}), 
lead to the well known reaction equations, which in non-dimensional form
read as
\qba
\dot c_1 &=& c_1^2c_2 -(1+b) c_1 \lb{eRea}\\
\dot c_2 &=& bc_1 -c_1^2c_2\nbc
\qea
Here $c_1$ and $c_2$ are the concentrations of constituent $X$ and $Y$
respectively. The parameter
$b$ presents the ratio between the decay rate of $X$ into harmless
substances and the decay rate of $X$ into pollutant $Y$. 

Without a source term the concentrations tend to zero. In our model
we assume for simplicity a point like pollutant source that is fixed, 
but encounters the fluid field of a two-dimensional isolated vortex.
This vortex produces a circular flow with velocity
of modulus $\Gamma/r$
at distance $r$ from the center, $\Gamma$
being the strength of the vortex. 
This problem is one-dimensional since the chemical reactions can only
take place along the circular streamline containing the source
and can be mapped to the unit interval with periodic boundary
conditions. Thus the flow is completely represented by one parameter,
the dimensionless velocity $v$.
We obtain the following combined reaction transport equations:
\qba
\frac{\partial }{\partial t}c_1 &=&s\delta (x) +
   c_1^2c_2 -(1+b) c_1 -v\frac{\partial }{\partial x}c_1 +
 \epsilon \frac{\partial^2}{\partial x^2}c_1 \lb{reaRest}\\		
\frac{\partial }{\partial t}c_2 &=&bc_1 -c_1^2c_2 - 
v\frac{\partial }{\partial x}c_2 +
\epsilon \frac{\partial^2}{\partial x^2}c_2\nbc
\qea
where $s$ is the strength of the source located
at $x=0$.
We allow here for a very small amount of diffusion with diffusion constant
$\epsilon $.

 Apart from the diffusion constant $\epsilon$ the equations contain
three relevant parameters, the decay ratio $b$, the source strength $s$
and the velocity $v$.

Eq.(\rep{reaRest}) represent the Eulerian description of an observer at rest. 
If we change to
the Lagrangian description (i.e. to an observer 
moving with the fluid) 
eq.(\rep{reaRest}) is transformed into
\qba
\frac{\partial }{\partial t}c_1 &=&s
\sum_{n =0}^{\infty} \delta (\bar x+vt+n) +c_1^2c_2 -(1+b) c_1
+\epsilon \frac{\partial^2}{\partial \bar x^2}c_1 \lb{reaMov}\\
\frac{\partial }{\partial t}c_2 &=&bc_1 -c_1^2c_2 +
\epsilon \frac{\partial^2}{\partial \bar x^2}c_2\nb
\bar{x} &=& x-vt\; mod\, 1\nbc
\qea
The solutions to this model will be discussed in the following 
two sections.

\section{PROPERTIES OF THE MODEL WITHOUT DIFFUSION}
\label{sec3}

First let us discuss the transport reaction equations, eq.(\rep{reaRest}) and
eq.(\rep{reaMov}) 
with diffusion switched off:\\
For $\epsilon =0$ the coordinate $\bar{x}$ can be interpreted as a parameter
equivalent to a time translation $\bar x/v$ in the driving term.
Therefore in the moving frame we have to solve for any $\bar x$
the {\em ordinary} differential equations
\qba
\frac{d}{dt}c_1 &=&s\sum_{n =0}^{\infty} \delta (\bar x+vt+n) +c_1^2c_2
-(1+b) c_1 \lb{reaOrd}\\
\frac{d}{dt}c_2 &=&bc_1 -c_1^2c_2 \nbc
\qea
The spatial distribution of the concentrations is completely determined
by the solution of the above equation and the initial distribution.
Obviously we have to take into consideration the time shift of
the driving for different points.

In eq.(\rep{reaOrd})
the difference to the 'usual' brusselator is the periodic $\delta $ 
function time 
dependence of the source. 
Thus we have a periodically driven (kicked) brusselator with one
extra parameter, the period of the driving $T \equiv 1/v$.
 The periodically driven brusselator has been investigated in different
contexts considering a constant plus a sinusoidal or delta function
time dependence of the source \cite{Sch88}\cite{KaiTom79}\cite{Aroetal86}.
In the $T \rightarrow 0$ limit the normal brusselator is recovered, i. e.
very frequent injections correspond to an almost uniform
source.
In this limit the parameter plane $s-b$ can be divided into two regions
(fig.1):
For values corresponding to higher source strengths the concentrations
converge to the fixed point $c_1^*=s, c_2^*=b/s$. As $s$ is decreased, the
fixed point becomes unstable and a Hopf bifurcation occurs along
the curve $s=\sqrt{b+1}$ forming the boundary between the two regions.
Below this curve the system converges to a limit cycle, i. e. the
concentrations oscillate periodically. (fig.2.)

As the periodic driving is switched on $0<T \ll 1$, a periodic pulsation
with period $T$ of the concentrations appears.
Moreover the initially two dimensional phase space becomes three
dimensional by including
the cyclic variable $t/T \; mod \; 1$ due to the
driving. Thus the dimensionality of the attractor increases as well,
and the original fixed point turns to a limit
cycle, representing a periodic time dependence, but still remains a
fixed point of the stroboscopic map.
Similarly, the original limit cycle becomes either a
torus corresponding to a quasiperiodic dynamics with one of the periods
equal to $T$ (fig.3) or a periodic orbit with large period. 
This is expected from the characteristic features
of periodically driven oscillators. There
exist resonant regions for driving frequencies close to their
natural frequency multiplied by a rational number. These resonant regions
appear here
below the Hopf bifurcation curve. In the parameter space of these 
regions Arnold tongues are detected analogous to those
of the so called
'circle map'\cite{Ott93} (fig.4,5). 
 As another effect of the driving the Hopf bifurcation curve moves to
smaller values of $s$ as $T$ increases (fig.1).
Since the dynamics is given by a set of two non-autonomous ordinary
differential
equations chaotic behavior is also possible for some values of the 
parameters leading to a strange attractor in the stroboscopic section
(fig.6).

Next we consider the {\em distribution} of the concentrations due to the
properties of the driven system.
Let us discuss the typical cases:\\
$\alpha $){\em  periodic time dependence with period $T$}:\\
This behavior occurs for large 
$T$, i.e. as the velocity of the flow becomes small.
In this case the concentrations oscillate 
and the phase of the oscillations is 
determined by the phase of the driving $t/T \;mod\; 1$.
Thus the final state does not depend on
the initial conditions. 
 In the moving frame
the only difference in the periodic time dependence at different points
of the flow is a time lag $\bar xT$.
\qb 
c(\bar x,t)=c(0,t+\bar xT).
\qe
If we move back to the standing frame ($\bar x \rightarrow x-vt$)
we obtain asymptotically a stationary state
$c(x,t)=c(0,xT)$ 
i.e. the concentrations converge to a 
$x$-dependent function
constant in time.

\noindent 
$\beta$) {\em  periodic oscillations with period $nT$}:\\
This behavior corresponds to the resonant regions.
In this case the concentrations can take $n$ different values for
a given phase of the driving depending on the initial conditions. 
\qb
c(\bar x,t)=\{ c(0,t+iT+\bar xT) \},\;\;i= 0...n-1,\lb{nPer}
\qe
The boundary between the basins of attraction of the $n$ branches 
of the solution is
a twisted (Mobius-like) surface so that the basin of attraction
of branch $i$ becomes the basin of attraction of branch
$i+1$ mod n after one period $T$.
Thus any smooth initial condition must have at least one
intersection with this surface. At this point the concentrations
converge to two different branches so a discontinuity appears
in the spatial distribution of the concentrations (fig.7). 
Note that this step is not a consequence 
of the
delta function in eq.(\rep{reaOrd}) but is due to geometrical constraints.
An initially random distribution can lead to a completely staggered
distribution whose envelopes are the $n$ branches of the solution.
In the moving frame the steps remain at the same position $\bar x$.

\noindent 
$\gamma$) {\em  quasiperiodic time-dependence}:\\
This is present in a region below the Hopf bifurcation curve
between the resonances and is pronounced for small $T$ because with 
increasing $T$ the size of the region below the Hopf curve
shrinks and at the same time the resonant islands grow in size.
This case corresponds to a motion on a torus in the 
phase space. The dynamics can be characterized by two cyclic angle-like
variables, one of them is the phase of
the driving and the other one depends
smoothly on the initial
concentrations.
$c(\bar x,t)= c(\bar x,t+\tau(\bar x))=c(t+\tau(\bar x)+\bar xT)$
Therefore an initially smooth distribution remains smooth in $\bar x$
for all times (except at the initial position of the source where the
time lag of the driving by $T$ leads to a discontinuity.)

$\delta)$ {\em  chaotic time-dependence}:\\
In this case the time dependence is very sensitive to the initial
conditions and thus the
distribution becomes irregular
on each scale regardless 
how smooth the initial distribution
may have been (fig.9a).

\section{THE ROLE OF DIFFUSION} 
\label{sec4}

Without diffusion the final distributions (except
those with the period $T$) 
have infinite 
degeneracy due to an arbitrary uneven\footnote{ we do not count the strong
increase of the $c_1$ concentration due to the $\delta $ function shape
as a step.}
 number of steps. 
Therefore diffusion is 
a {\em singular} perturbation that has significant consequences for the
system as small as $\epsilon $ may be. For the following computations
we used the Crank-Nicholson scheme combined with operator 
splitting \cite{Preetal92}.

We discuss here 
the simplest nontrivial case of eq.(\rep{nPer}) first, which occurs
for period $2T$. We denote with a $-+$ ($+-$)
step an 'upward' ('downward') steep increase (decrease)
of the concentration, but exclude 
the strong increase of $c_1$ at the location of the source.
If the diffusion is small enough we can treat a step as 
isolated (for a very long time). Due to diffusion the step
will
move with a drift velocity, cf (fig.7).
Scaling and symmetry
arguments suggest that it should be proportional to a higher power of
$\sqrt{\epsilon }$ and in fact we find numerically
a dependence $\propto \epsilon $. The important point, however, 
is that, averaged
over $2T$, each isolated step moves with the {\em same} drift velocity. 
In
fact after time $T$  a $-+$ step becomes a $+-$ step  and vice versa.
What we expect then as the essential ingredient of eq.(\rep{nPer}) is that $f$
tries to enforce solutions with period of $2T$. In appendix A we
have derived a simple function ${\bf f}$ that has just this property {\em and}
makes it possible to treat eq.(\rep{nPer}) analytically. Then we find: 
first, for times 
\qb
T_1 =\mathcal{ O}(1)
\qe
all steps
with distance of $\mathcal{ O}(\sqrt{\epsilon })$ vanish. 
During this time the diffusion
does nothing but a coarse graining. Second, over a period of about
\qb
T_2 = e^{\beta/\sqrt{\epsilon }},\; \beta  =\mathcal{ O}(1)
\qe
all other steps are effected. The diffusion removes the degeneracies by and
by, until a final state
emerges that has no steps at all, besides the generic one that cannot be
removed. This state has global stability in our model.
Numerically we find the same phenomena for ${\bf f}$ of the brusselator, 
cf. fig.7b. 

The effects described here occur quite 
independently from how small $\epsilon $ is again demonstrating that
the diffusion is a singular
perturbation. On the other hand $T_2$
depends exponentially on $1/\sqrt{\epsilon }$. When a further 
perturbation has to be added acting on a time scale $\tau $ we expect 
quite different situations 
depending on whether $T_2 >\tau $ or $T_2 <\tau $. This means that
the effect of such a perturbation depends {\em sensitively} on 
$\sqrt{\epsilon} $
which shows that introducing diffusion as a parameter - as has been
done by introducing 'turbulent diffusion' - is quite a dangerous
approximation.

We expect even more complicated properties of the concentrations
having higher periods (in absence of diffusion). There are
two reasons for that: 
i) if the period is $nT$ the system has at any location
$n-1$ choices for the height of a step, ii) the steps are no longer
equivalent but are separated in classes and only steps within the
{\em same} class change into each other and therefore
move with the same mean drift velocity $v_d$.
Indeed, the effect of diffusion on the periodic solution can be very
significant in some cases by leading to a complicated irregular behavior
of the system in space and time. As can be seen from fig.8,
inside the chaotic concentration field coherent regions with regular periodic
time dependence appear and disappear continuously. This kind of spatiotemporal
intermittency has been observed in different extended 
systems\cite{Kan84})\cite{ChaMa88})
(e.g. in case of coupled maps\cite{KiHla86})\cite{SchMa82}\cite{OPRK97}.
If one starts the 
simulation with a smooth initial distribution first at least the intrinsic
step appears as described above. The pertubation of the periodic solution
around the discontinuity leads to a chaotic time dependence which 
due to the diffusive coupling spreads 
over the whole system. Such behavior can be
observed for parameters which lie
in the vicinity of the chaotic regimes in the $\epsilon =0$ case. 
The solution appears already
for very small $\epsilon $ demonstrating again that diffusion is a singular
perturbation.

In case of quasiperiodic local behavior, instead of a finite number
of discrete branches, a continuous set of solutions exists filling
the torus in the phase space. Thus the discontinuity present in the
case without diffusion is easily removed by an arbitrarily weak diffusion
leading to coherent quasiperiodic oscillations of the whole system.

When the parameters correspond to chaotic local dynamics,
diffusion tends to form correlated regions of finite extent in space and time
(fig.9) 
similar to the case above.
As $\epsilon$ is increased, the local dynamics becomes completely regular
with a frozen irregular distribution in space which certainly depends on 
the initial distribution. 

\section{Conclusion}
\label{concl}

High peaks can appear in periodic solutions of 
chemical reaction equations in which the constituents are {\em homogeneously}
mixed tracer gases of the atmosphere. However, depending on the motion
of the fluid the mixing need not be homogeneous
at all, and the
question arises how these solutions will then change. 

In this paper we investigate this question for a simple model,
the brusselator with pointlike source in a one-vortex flow.
Simple as the model appears, it already
demonstrates the strong modifications occurring
as soon as we move away from the homogeneous situation. One observes this
when computing the concentration distribution along the (closed)
streamline
in which the source is located. As a function of time
we detect solutions that are very similar to those 
of the homogeneous case. This happens as long as the period $T$ 
of the flow 
is small. Furthermore we find 
solutions with period $nT$, moreover quasiperiodic and chaotic ones.
All these solutions, except that with period $T$ are infinitely
degenerate and therefore depend on the initial
distribution. Even if that is smooth, the distributions can have 
asymptotically an arbitrary (uneven)
number of discontinuities, in the chaotic case on {\em each scale}.

In such situations diffusion is a singular perturbation and 
switching on arbitrary small 
diffusion along the streamline has two effects: First 
after a time of $\mathcal{ O}(1)$
it leads to a 'coarse graining' of the distribution on a
space scale $\propto \sqrt{\epsilon }$
where $\epsilon $ is the strength of the diffusion. Second on a time scale 
$\propto e^{\sqrt{\alpha} /\sqrt{\epsilon }}$, 
$\alpha  =\mathcal{ O}(1)$ it removes all 
discontinuities but one for
solutions which have (without diffusion) period $2T$. This shows that
the solutions depend sensitively on $\sqrt{\epsilon }$.
For parameters that lead
(without diffusion) to solutions of higher period, quasiperiodic solutions
and chaotic ones
were observed, moreover
pattern formation and spatio temporal chaos. All those solutions have nothing
in common with the case of homogeneous mixing we started with.  

Although this one-dimensional model is far from being a 
realistic representation
of the chemistry and transport in the atmosphere, it shows that even a trivial
non-turbulent flow interacting with a simple regular chemical dynamics of
just two reactants can lead to a complex irregular behavior of the
concentration fields. 

acknowledgments
This  work has been supported in part by the 
German - Hungarian Scientific and Technological Cooperation 
{\em classical and quantum chaos and applications}.
One of us (Z.N.) would like to thank the group of the modeling forum
for their kind hospitality at the research center J\"ulich where 
part of this work
had been done. We thank Gert Eilenberger and
Tam\'as T\'el for useful discussions.

\newpage

\vspace{7mm}
\begin{appendix}
\section{appendix}
\label{appa}

In this appendix we derive the properties of eq.(\rep{startEq}) 
for our model assuming that without diffusion the solution has a
$2T$ period in the moving system. We have
\qb
\partial_t {\bf c}= {\bf f}({\bf c},\bar{x},t) +\epsilon \partial^2_{\bar{x}} 
{\bf c}\lb{orig}
\qe
with the periodic boundary conditions
\qba
{\bf c}(\bar{x},t) &=&{\bf c}(\bar{x}+1,t)\\
\partial_{\bar{x}}{\bf c}(\bar{x},t) &=&
\partial_{\bar{x}}{\bf c}(\bar{x}+1,t)\nbc
\qea
This equation holds true in the frame moving with a velocity
$v =1/T$.

The use of ${\bf c}$ can be awkward since the components have to
be positive. Therefore we write
\qb
{\bf n} ={\bf c} +{\bf const}
\qe
and get the equation for ${\bf n}$
\qba
\partial_t {\bf n}&=& {\bf g}({\bf n},\bar{x},t) +
\epsilon \partial^2_{\bar{x}} {\bf n}\nb
&with&\lb{transfEq}\\
{\bf g}({\bf n},\bar{x},t) &=&{\bf f}({\bf const}+{\bf n},\bar{x},t)\nbc
\qea
and the periodic boundary conditions
\qba
{\bf n}(\bar{x},t) &=&{\bf n}(\bar{x}+1,t)\lb{boundn}\\
\partial_{\bar{x}}{\bf n}(\bar{x},t) &=&
\partial_{\bar{x}}{\bf n}(\bar{x}+1,t)\nbc
\qea

Without diffusion 
${\bf n}$ moves exponentially fast to its 
asymptotic limit ${\bf n}^{(0)}$ having the properties
\qba
{\bf n}^{(0)}(\bar{x},t) &=&{\bf n}^{(0)}(\bar{x},t+2T)\nb
&\mbox{and either}&\nb
{\bf n}^{(0)}(\bar{x},t) &=&{\bf n}^{(0)}(t -\bar{x}T)\lb{condn0}\\
&or&\nb
{\bf n}^{(0)}(\bar{x},t) &=&{\bf n}^{(0)}(T+t -\bar{x}T)\nbc
\qea
$n^{(0)}$ and its properties remain important also if diffusion is
switched on since $g$ can be expanded around $n^{(0)}$.

To understand the physics of 
eq.(\rep{transfEq}) with conditions eq.(\rep{condn0}) 
we construct a simple 
model
for the function $g$
in three steps:\\
\\
\underline{step 1:}
we introduce a very simple ${\bf n}^{(0)}$
\qba
n_1^{(0)}(\bar{x},t) &=&\Re\{ae^{i\pi (t/T-\bar{x})}\}\lb{simp_n0}\\
n_2^{(0)}(\bar{x},t) &=&\Im\{ae^{i\pi (t/T-\bar{x})}\}\nbc
\qea
and we represent the two dimensional vectors by complex
numbers.\\
\\
\underline{step 2:}
we use the ansatz
\qb
n(\bar{x},t)=e^{i\pi (t/T-\bar{x})}\cdot m(\bar{x},t)\lb{ansatzn}
\qe
and get the pde for $m$
\qba
\partial_tm &=&\tilde{g} -i\pi \epsilon \partial_{\bar{x}}m -
\pi^2\epsilon m +\epsilon \partial_{\bar{x}}^2m\nb
&with&\\
\tilde{g} &=&e^{-i\pi (t/T-\bar{x})}
g(e^{i\pi (t/T-\bar{x})}m,\bar{x},t) -(i\pi/T)m \nbc
\qea
The terms $i\pi \epsilon \partial_{\bar{x}}m$ and 
$\pi^2\epsilon m$ are of higher order in $\sqrt{\epsilon }$
and will be left out for simplicity.
The boundary conditions of eq.(\rep{boundn}) are replaced by
\qba
m(\bar{x},t) &=&-m(\bar{x}+1,t)\lb{boundm}\\
\partial_{\bar{x}} m(\bar{x},t) &=&-\partial_{\bar{x}} m(\bar{x}+1,t)\nbc
\qea
\\
\underline{step 3:}
we construct a simple $\tilde{g}$. Because of eq.(\rep{simp_n0}) and
eq.(\rep{ansatzn}) $m^{(0)}$ can take two values only,
\qb
m^{(0)} = \pm a
\qe
and $a$ can be chosen to be real and positive. When $m$ is in the 
neighborhood of $m^{(0)}$ $\tilde{g}$ can be expanded and we obtain
\qba
\tilde{g} &=&-\alpha (\bar{x},t)(m -a) + ...\nb
&or&\\
\tilde{g} &=&-\alpha (\bar{x},t)(m +a) + ...\nbc
\qea
For $\alpha (\bar{x},t)$ we insert a real 
positive constant\footnote{$\alpha $ could be a complex constant
as well as long as the real part is positive.}. The linear approximation
of $\tilde{g}$ is of course incorrect if $m$ is not close to $\pm a$.
A nonlinearity is added simply by the prescription
$$
\tilde{g} = \left\{\begin{array}{l}
             -\alpha (m -a) \mbox{ for $\mid m-a\mid  <\mid m+a\mid $}\\
             -\alpha (m +a) \mbox{ else}\end{array} \right.\\
$$

Thus we get the partial differential equation:

\qba
\partial_tm &=&\tilde{g} +\epsilon \partial_{\bar{x}}^2m\nb
&with&\lb{modelEq}\\
\tilde{g} &=& \left\{\begin{array}{l}
             -\alpha (m -a) \mbox{ for $\Re{m} >0$}\\
             -\alpha (m +a) \mbox{ for $\Re{m} \leq 0$}\end{array} \right.\nbc
\qea
Boundary conditions are given by eq.(\rep{boundm}). 
The connection between n and $m$ 
is given by eq.(\rep{ansatzn}) and real and imaginary part
of $n$ are the components of ${\bf n}$.

\vspace{17mm}
\centerline{{\bf properties of the solutions of} eq.(\rep{modelEq})}

\vspace{7mm}\noindent 
I) Diffusion switched off, i.e. $\epsilon =0$,

$m$ consists asymptotically of
an uneven number of steps with values $\pm a$. The number of steps
can be arbitrarily high and
is determined exclusively by the initial distribution of $m$.

\vspace{7mm}\noindent 
II) Diffusion switched on, i.e. $\epsilon >0$, 

\vspace{7mm}\noindent 
1) Isolated step\\

we assume that there is a constant velocity $\sqrt{\epsilon }w$ with which the
step is moving. Transforming to new coordinates $y$ with
\qb
\bar{x} =y +\sqrt{\epsilon }w t
\qe
and assuming that the step occurs at $y=0$ the two equations
are to be solved:
\qbas	
0 &=&-\alpha (m+a) +\sqrt{\epsilon }w\partial_ym +\epsilon \partial_y^2m,\; 
y <0\\
0 &=&-\alpha (m-a) +\sqrt{\epsilon }w\partial_ym +\epsilon \partial_y^2m,\; 
y >0
\qeas
because of the boundary conditions for the
isolated step
\qbas
m(-\infty ) &=& -a\\
m(\infty ) &=& a
\qeas
the solution is
\qba
m_-(y) &=&Ae^{\gamma y/\sqrt{\epsilon }} -a,\; 
\gamma  =(1/2)(-w +\sqrt{4\alpha  +w^2}),\; y \leq 0\lb{gaqEq}\\
m_+(y) &=&Be^{\tilde{\gamma }y/\sqrt{\epsilon }} +a,\; 
\tilde{\gamma } =(1/2)(-w -\sqrt{4\alpha  +w^2}),\; y \geq 0\nbc
\qea
with the boundary condition
\qbas
m_-(0) &=& m_+(0)\\
m_-'(0) &=& m_+'(0)\\
\qeas
Because of eq.(\rep{modelEq}) there is the further condition
$$
\Re\{m_-(0)\} =0
$$
Therefore $A =a$ and $B =-a$, $m$ is real and the condition for $w$ is
obtained from
$$
a\gamma  =-a\tilde{\gamma}
$$
which means
$$
w=0
$$

\vspace{7mm}\noindent 
2) two interacting steps isolated from the rest\\

Let the $-+$ step be left, the $+-$ step be right. Both steps move
because of interacting with each other and we assume that
the interaction changes speed and shape of the steps only slowly 
(the distance $2x_d$ between them decreases of course).

First we rescale to avoid the $\epsilon $ dependence
\qb
\xi =\frac{\bar{x}}{\sqrt{\epsilon }}
\qe
Next we transform into a coordinate system moving with the $-+$ step, whose
position is at $0$. we obtain 
\qb
\xi =\eta  +w_dt
\qe
and
\qba 
0&=&-\alpha (m_- +a) +w_d
\partial_\eta m_- +
\partial^2_\eta m_- , \;\eta \leq 0 \lb{pdeMov} \\
0&=&-\alpha (m_+ -a) +w_d
\partial_\eta m_+ +
\partial^2_\eta m_+ , \;\eta  \geq 0 \nbc
\qea
Boundary conditions:
\qba
m_-(-\infty )  &=& -a\nb
m_-(0)  &=& m_+(0)\lb{bCond}\\
m_-'(0) &=& m_+'(0)\nbc
\qea
Furthermore the presence of the $+-$ step is taken care of by the condition  
\qb
m_+'(\xi_d)    = 0 \lb{vEq}
\qe
and we have the constraint (cf eq.(\rep{modelEq}))
\qb
\Re\{m_-(0)\}    = 0\lb{conW}
\qe
Then we get with an exponential ansatz (cf eq.(\rep{gaqEq})
\qba
m_-(\eta ) &=& A e^{\gamma \eta } -a,\; \eta  \leq 0,\\
m_+(\eta ) &=& B e^{\tilde{\gamma }\eta } +C e^{\gamma \eta }+a,\; 
0\leq \eta  \leq \xi_d\lb{sol}\nbc
\qea
Again $m$ can be chosen to be real and
the conditions eq.(\rep{bCond}), eq.(\rep{vEq}) and eq.(\rep{conW}) yield
\qba
A-a &=& B +C +a\nb
A\gamma  &=& B\tilde{\gamma } +C\gamma \\
0 &=& B\tilde{\gamma }e^{\tilde{\gamma }\xi_d} +C\gamma e^{\gamma \xi_d}\nb
A-a &=& 0\nbc
\qea
From these equations we get $w_d$ (neglecting all 
terms $w_d^2$ and
higher)   
\qb
w_d \approx 2\sqrt{\alpha }e^{-2\sqrt{\alpha }\xi_d}\lb{w_dEq}
\qe
which is correct for
\qb
\sqrt{\alpha }\xi_d >1\lb{res}
\qe

From scaling arguments we infer that eq.(\rep{res}) gives the correct
order of magnitude for $\sqrt{\alpha }\xi_d <1$.\\
One can use eq.(\rep{w_dEq}) also to prove that there is no stationary
state. If it were all the equations were exact and in particular
eq.(\rep{w_dEq})
which in turn would be a contradiction. 

Now we compute the lifetime $t_l$ of a step which is in the original
coordinates
\qb
t_l =\frac{1}{4\alpha }(e^{2\sqrt{\alpha }x_d/\sqrt{\epsilon }} -1)
\qe

\vspace{7mm}\noindent 
3) $n$ interacting steps\\

To treat this problem we take into account the interaction between
nearest neighbors only ( the interaction between next nearest
neighbors is 
exponentially small compared to the interaction between the nearest
neighbors). Then it is sufficient to look into the problem of one
step between two other steps.
We approximate the interaction again by boundary conditions and obtain
two conditions of the form eq.(\rep{vEq}). Doing an analogous 
computation with the same approximations
we obtain for the velocity of the step 
\qb
w_d \approx 2\sqrt{\alpha }
(e^{-2\sqrt{\alpha } x_{d+}/\sqrt{\epsilon }} 
-e^{-2\sqrt{\alpha }x_{d-}/\sqrt{\epsilon }})
\qe
Here $2x_{d+}$ ($2x_{d-}$) is the distance to the right (left) step. 
From this
result we conclude that all states with more than one step will be unstable
since two neighboring steps will annihilate each other.
\end{appendix}

\newpage

\begin{figure}[h]
\caption{
 Curves corresponding to the Hopf bifurcation in the parameter plane
$s-b$ for different values of the period $T$.}

\end{figure}

\begin{figure}[h]
\caption{ Constant in time and periodic behavior of the concentrations
$c_1$ and $c_2$ for the unforced brusselator ($T=0$). The
parameters are $s=2.5$, $b =3.0$ and $s=1.0$, $b =3.0$, respectively.}
\end{figure}

\begin{figure}[h]
\caption{ Quasiperiodic time dependence of the concentrations $c_1$
and $c_2$ (a), and stroboscopic section (b) for $s=1.0$, $b =3.0$ and
$T=1.0$.}\end{figure}

\begin{figure}[h]
\caption{ Stroboscopic plot of $c_2$ in function of $T$ for $s=1.9$ and
$b =7.7$. The Hopf bifurcation occurs around $T=1.03$ and there are
resonant windows 
inside the quasiperiodic region labelled by the ratio of the
two periods.}
\end{figure}

\begin{figure}[h]
\caption{ Periodic(blank) and quasiperiodic(gray) regions in a section
of the parameter space for $T=1.0$. The behavior of the system was
identified by calculating the leading Lyapunov exponent which is smaller
then $-0.0025$ for the blank region.}\end{figure}  

\begin{figure}[h]
\caption{ Chaotic time dependence of the concentrations $c_1$ and $c_2$
and the stroboscopic section of the strange attractor. The parameters
are $s=1.2$, $b =7.0$ and $T=1.36$.}\end{figure}

\begin{figure}[h]
\caption{ Spatiotemporal plot of the concentrations $c_1$ (left) and $c_2$
(right) along the streamlines in the co-moving frame 
represented on a grayscale, so that 
concentrations increase from
black to white.  
The simulation was
started with both concentrations equal to zero and the initial position
of the source is at $x=0.2$. Parameters are $s=1.0$ $b =5.0$ and $T=1.7$
In case (a) $\epsilon =0$ and a non-moving discontinuity is present at $x=0.2$.
When diffusion is switched on $\epsilon = 0.001$ the discontinuity becomes
rounded and moves (to the left in this case) along the streamline.}
\end{figure}

\begin{figure}[h]
\caption{ Stroboscopic spatiotemporal plot of concentration $c_2$
for parameters $s=0.8$, $b =6.0$ and $T=1.85$, that correspond to a 
periodic behavior
with period $3T$ when diffusion is neglected. Here 
$\epsilon =2 \cdot 10^{-5}$
that leads to an irregular spatiotemporal dynamics.}\end{figure} 

\begin{figure}[h]
\caption{ Stroboscopic spatiotemporal plots of concentration $c_2$
for $s=0.8$, $b =6.0$ and $T=1.89$. This parameters correspond to
a chaotic local dynamics when diffusion is not considered. We
assumed that the initial concentrations are randomly distributed
in a small interval $[0,0.0001]$ for both constituents. 
The diffusion coefficient is $\epsilon =0$ (a), 
$\epsilon =1.5 \cdot 10^{-5}$ (b) 
and $\epsilon =2 \cdot 10^{-5}$ (c), respectively.}
\end{figure}

\setlength{\unitlength}{1.0cm}

\newpage
\begin{picture}(26.5,10.7)

\put (-1.5,-3.0){
\epsfig{file =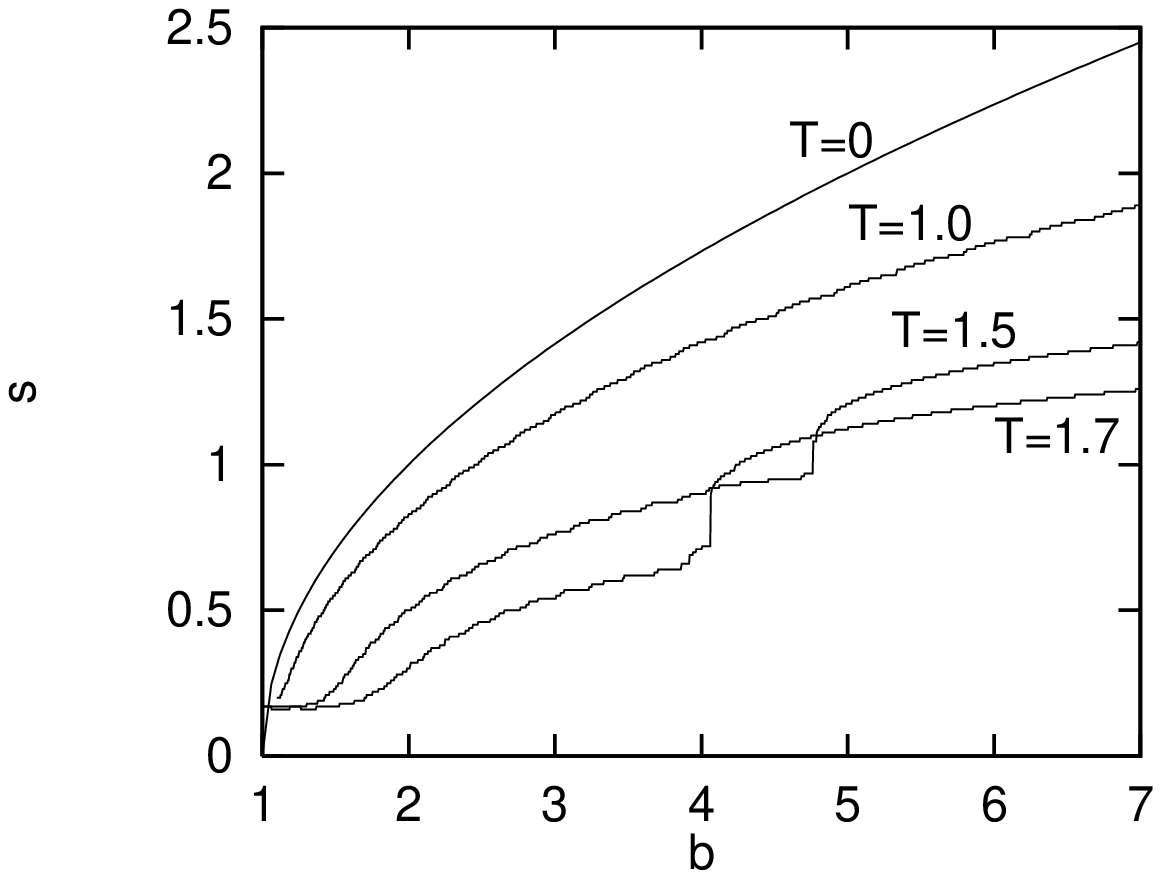,scale=1.0,,
  height =15cm,width =18cm,angle =00}}
\put(-4,-10.1){\makebox[8cm]{\centerline{\Large{fig.1}}}}
\end{picture}

\newpage
\begin{picture}(26.5,10.7)

\put (-.5,-3.0){
\epsfig{file =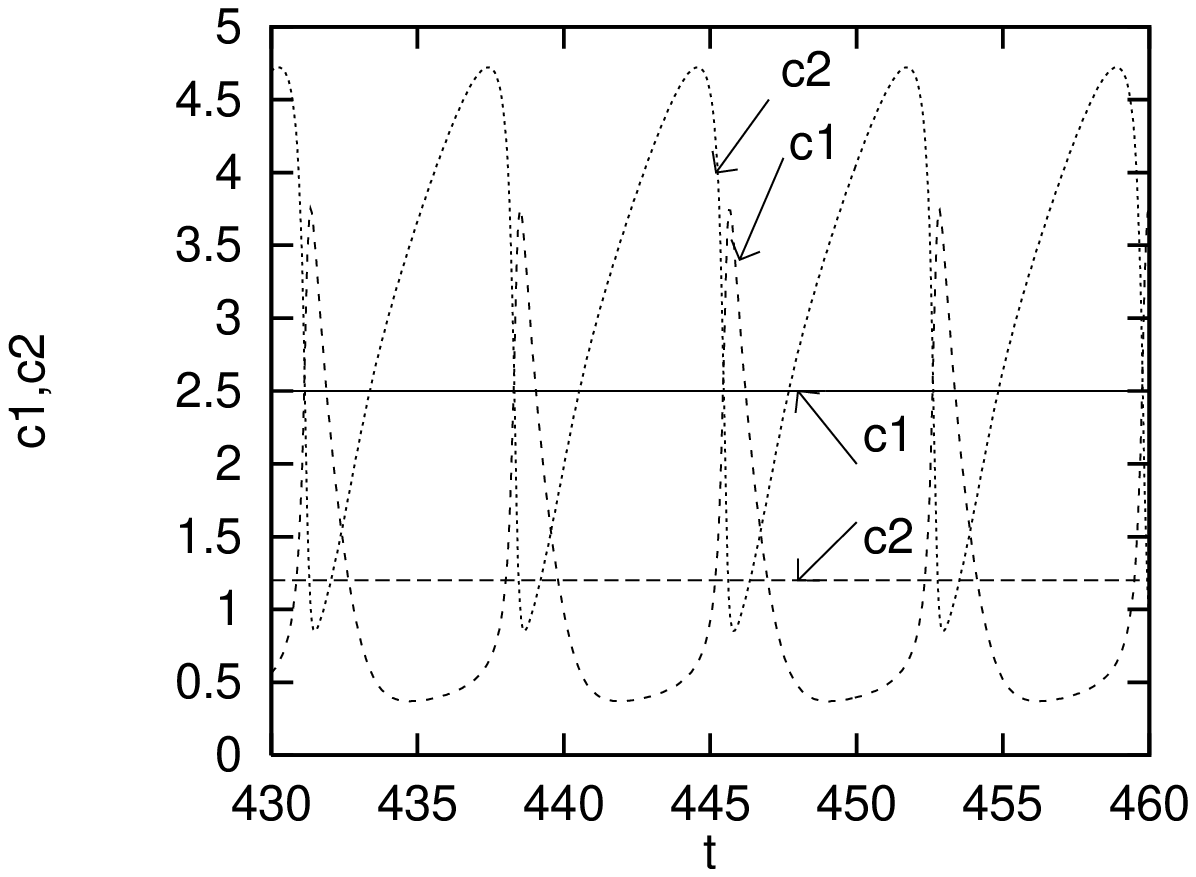,scale=1.0,,
  height =15cm,width =24cm,angle =00}}
\put(-4,-12.1){\makebox[8cm]{\centerline{\Large{fig.2}}}}
\end{picture}

\newpage
\begin{picture}(26.5,10.7)

\put (-.5,-13.0){
\epsfig{file =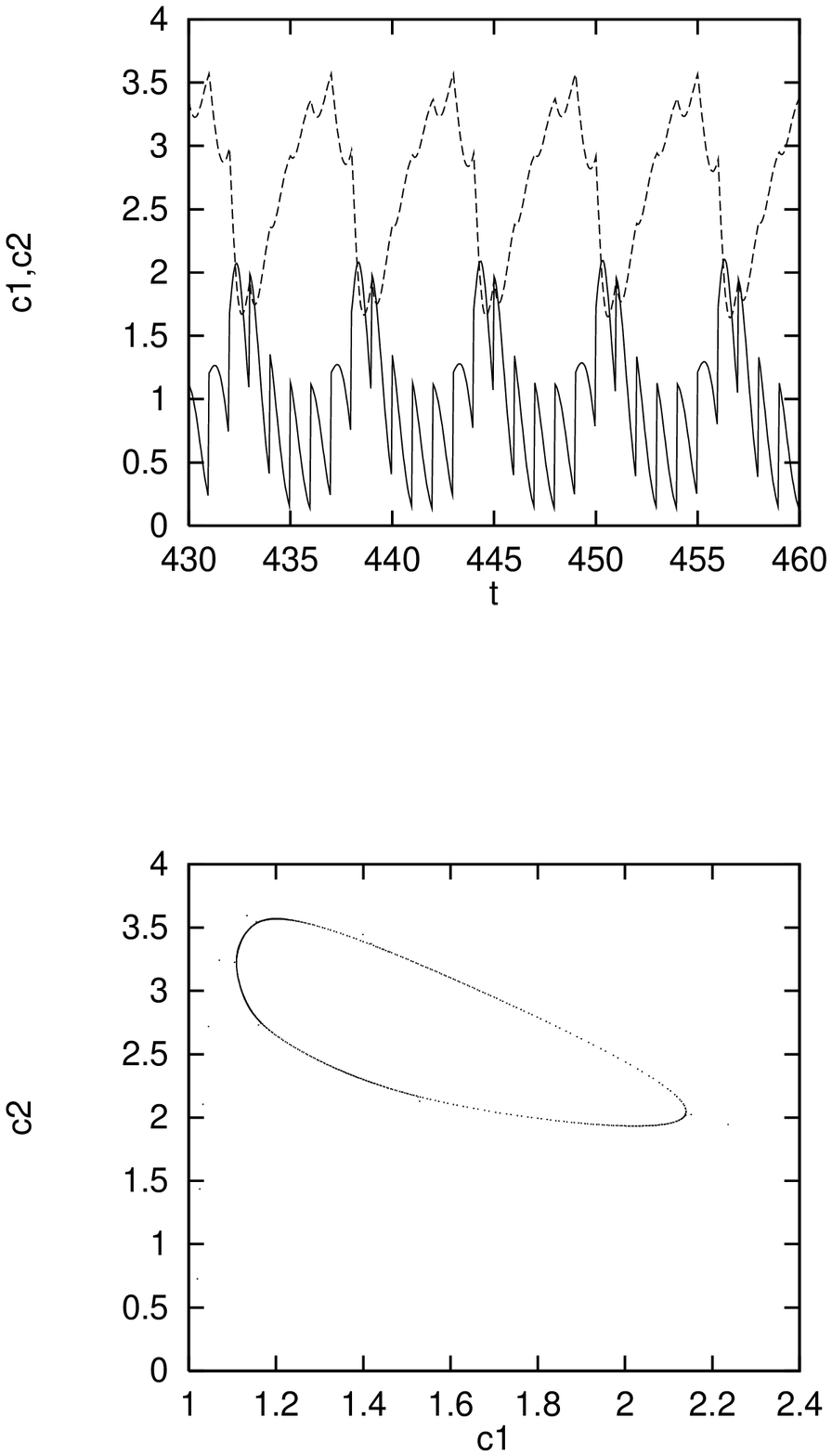,scale=1.0,,
  height =25cm,width =24cm,angle =00}}
\put(-4,-15.){\makebox[8cm]{\centerline{\Large{fig.3}}}}
\put(-5,3.){\makebox[8cm]{\centerline{\Large{a)}}}}
\put(-5,-11.){\makebox[8cm]{\centerline{\Large{b)}}}}
\end{picture}

\newpage
\begin{picture}(26.5,10.7)

\put (-.5,-3.0){
\epsfig{file =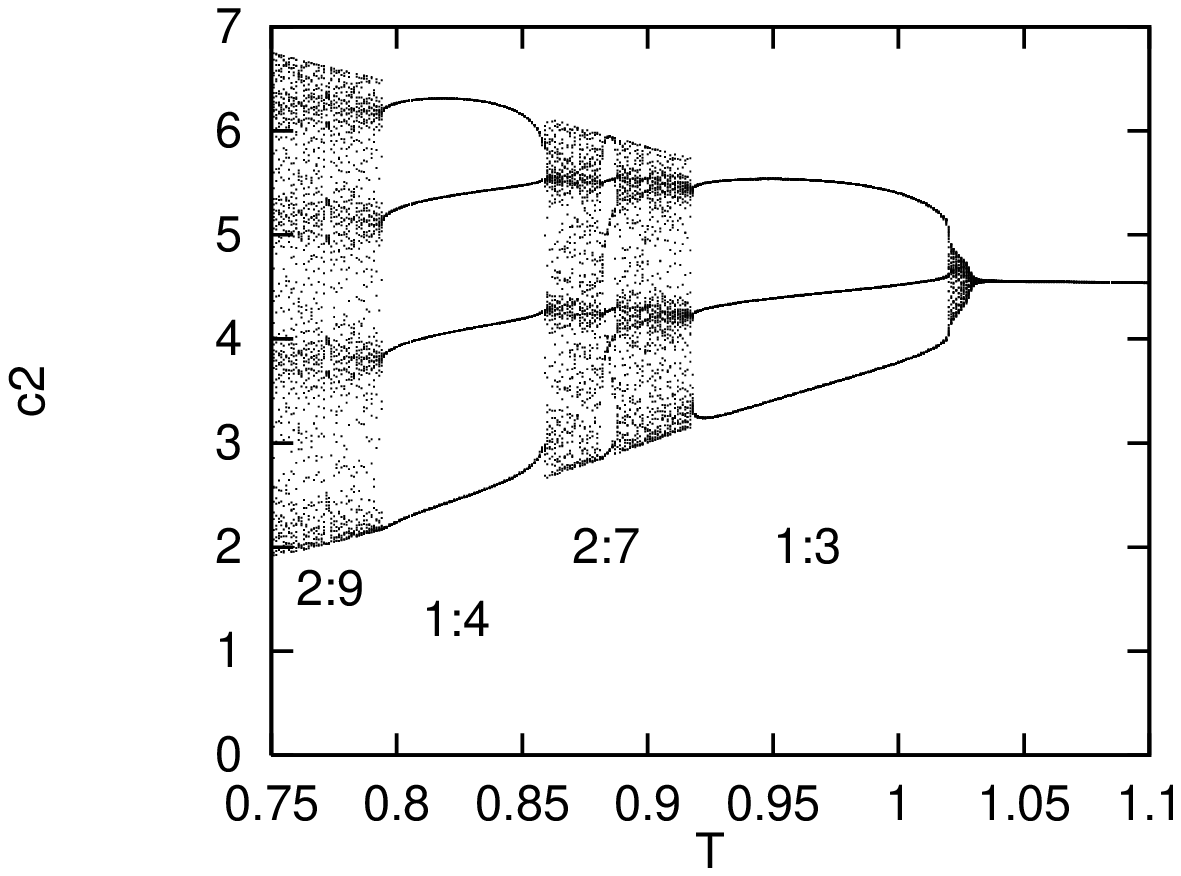,scale=1.0,,
  height =15cm,width =24cm,angle =00}}
\put(-4,-12.1){\makebox[8cm]{\centerline{\Large{fig.4}}}}
\end{picture}

\newpage
\begin{picture}(26.5,10.7)

\put (-.5,-3.0){
\epsfig{file =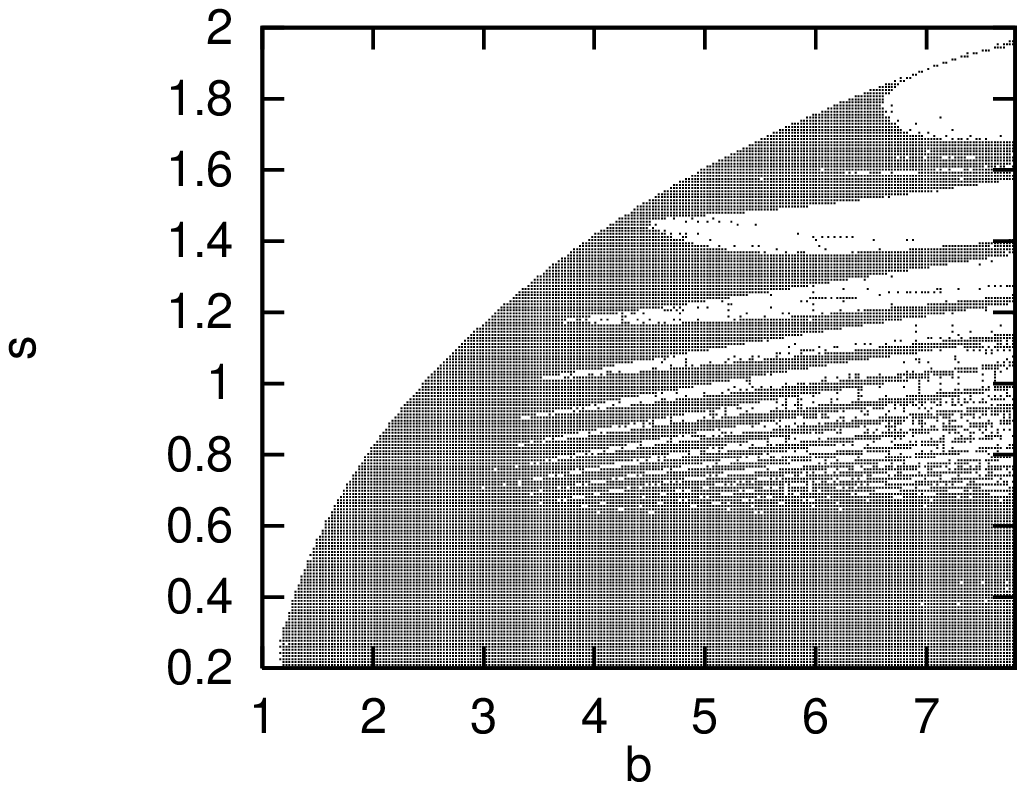,scale=1.0,,
  height =15cm,width =24cm,angle =00}}
\put(-4,-12.1){\makebox[8cm]{\centerline{\Large{fig.5}}}}
\end{picture}

\newpage
\begin{picture}(26.5,10.7)

\put (-.5,-13.0){
\epsfig{file =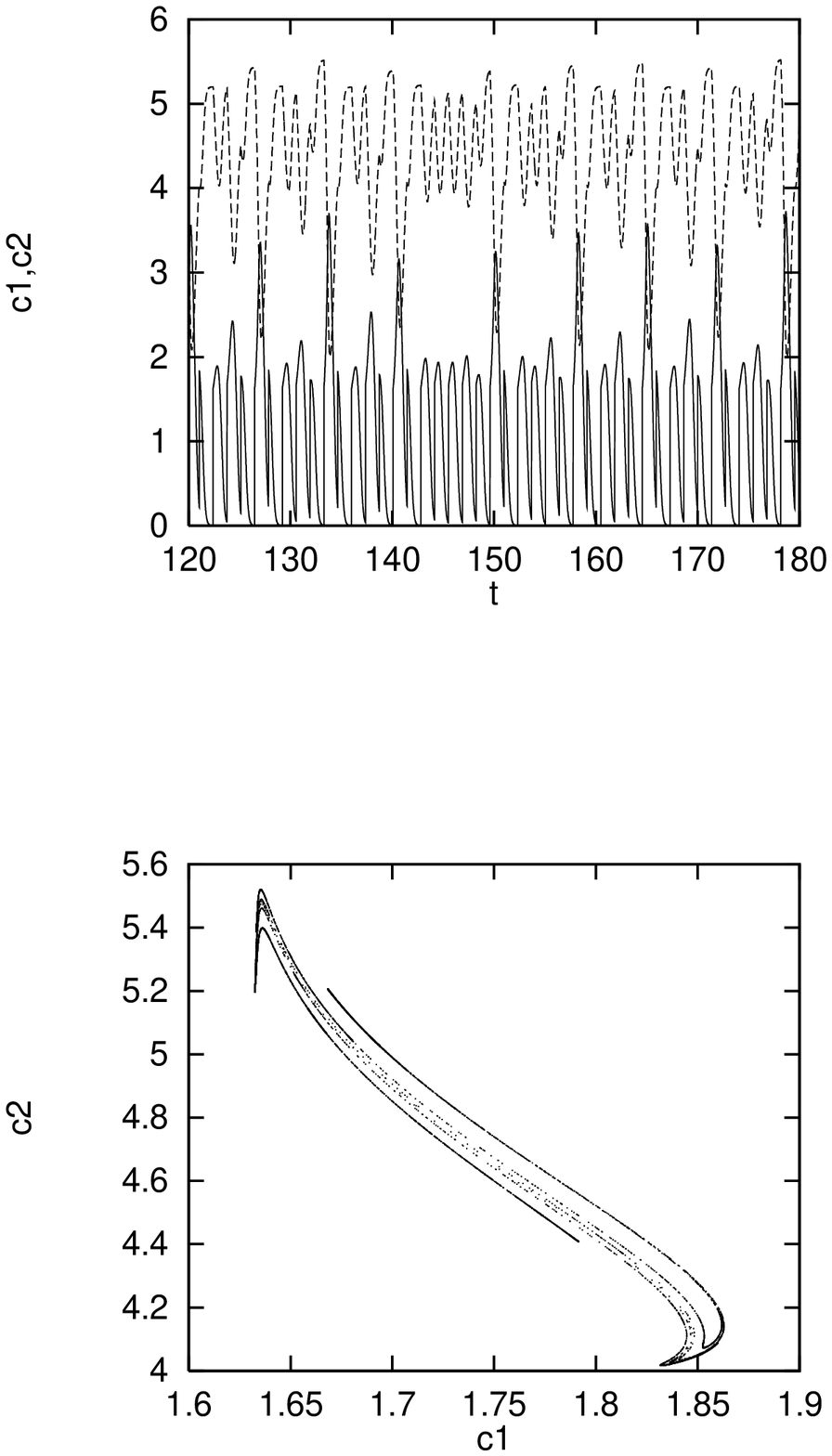,scale=1.0,,
  height =25cm,width =24cm,angle =00}}
\put(-4,-15.5){\makebox[8cm]{\centerline{\Large{fig.6}}}}
\put(-5,3.){\makebox[8cm]{\centerline{\Large{a)}}}}
\put(-5,-11.){\makebox[8cm]{\centerline{\Large{b)}}}}
\end{picture}

\newpage
\begin{picture}(26.5,10.7)

\put (1,-8.0){
\epsfig{file =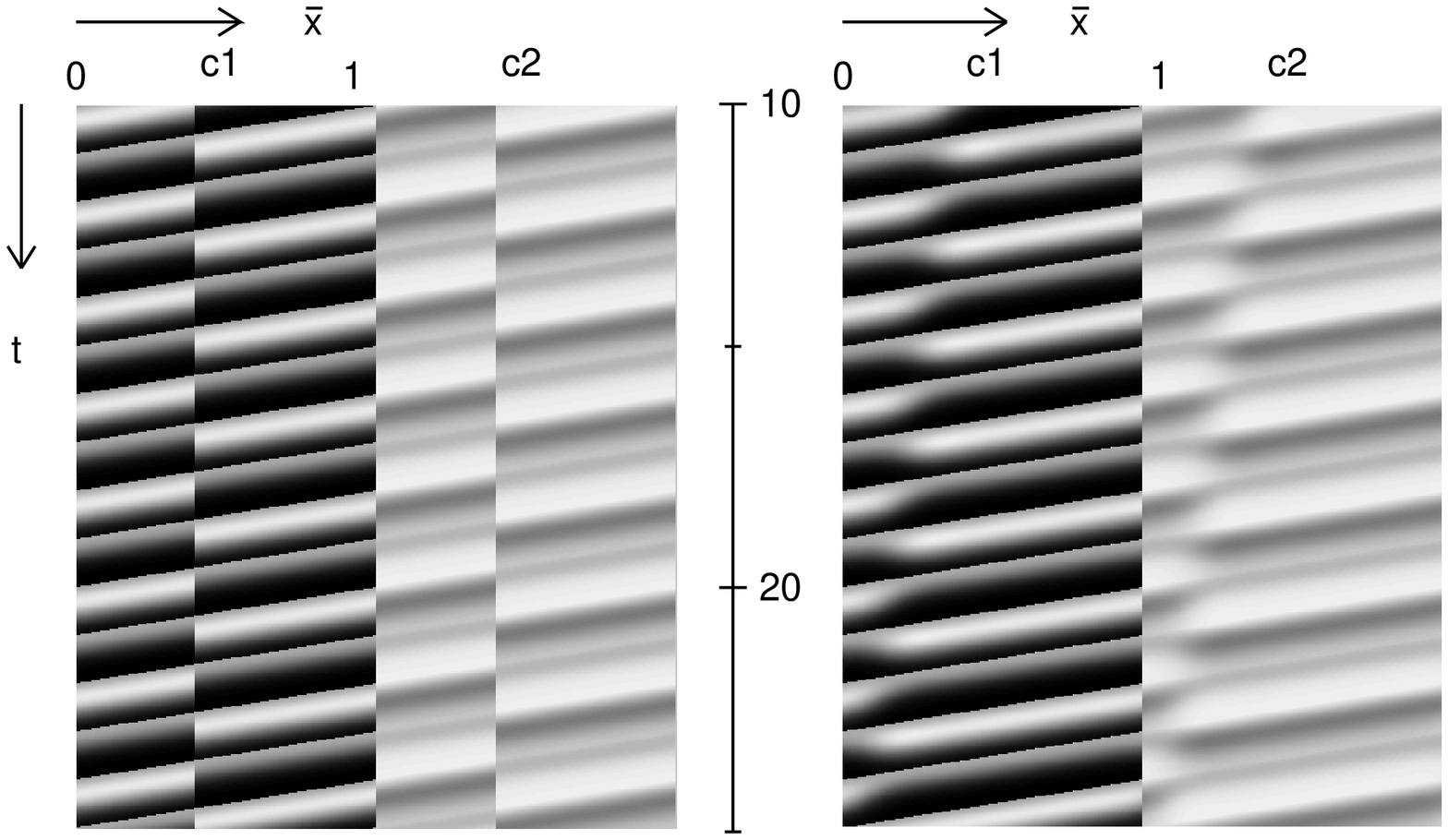,scale=1.0,,
  height =20cm,width =16cm,angle =00}}
\put(-4,-15.){\makebox[8cm]{\centerline{\Large{fig.7}}}}
\put(1,-10.){\makebox[8cm]{\centerline{\Large{a)}}}}
\put(9,-10.){\makebox[8cm]{\centerline{\Large{b)}}}}
\end{picture}

\newpage
\begin{picture}(26.5,10.7)

\put (2.5,-6.0){
\epsfig{file =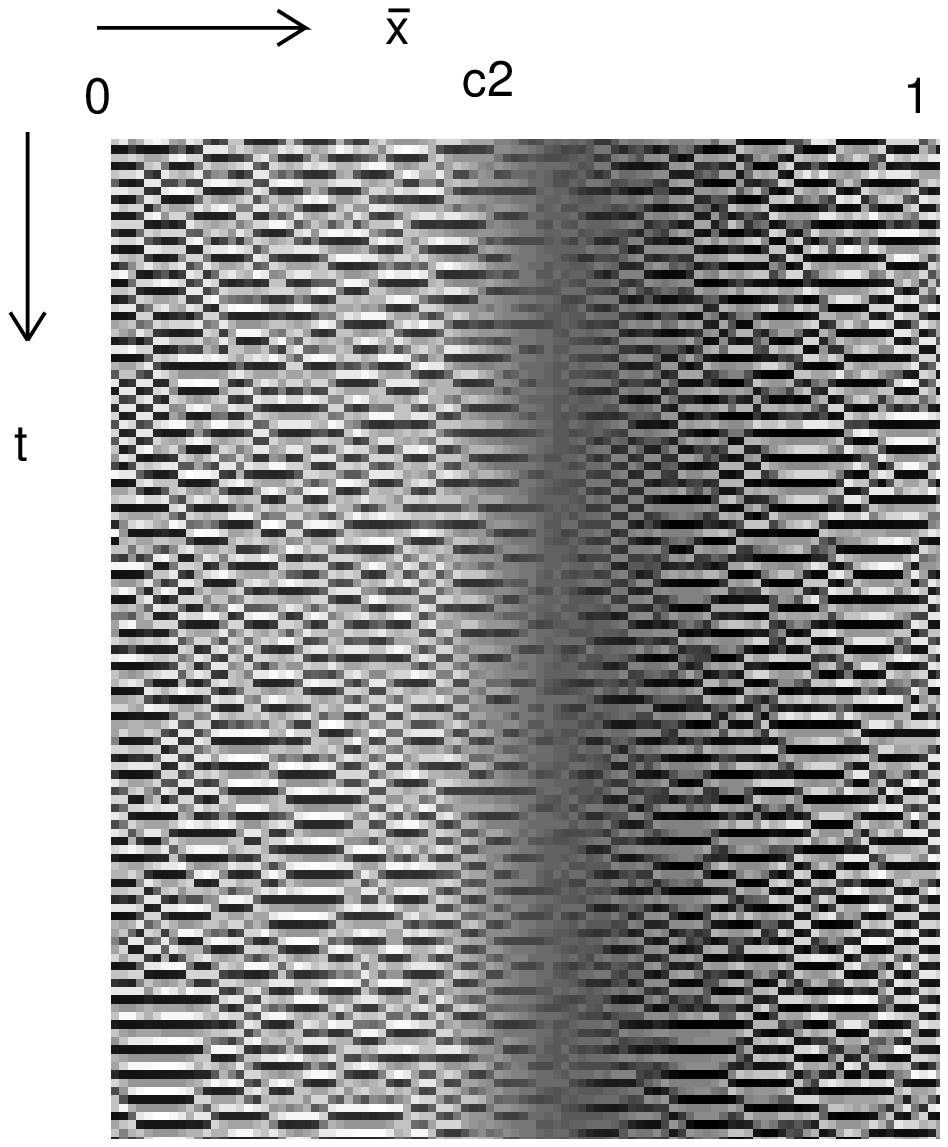,scale=1.0,,
  height =17cm,width =26cm,angle =00}}
\put(-4,-12.1){\makebox[8cm]{\centerline{\Large{fig.8}}}}
\end{picture}

\newpage
\begin{picture}(26.5,10.7)

\put (1,-8.0){
\epsfig{file =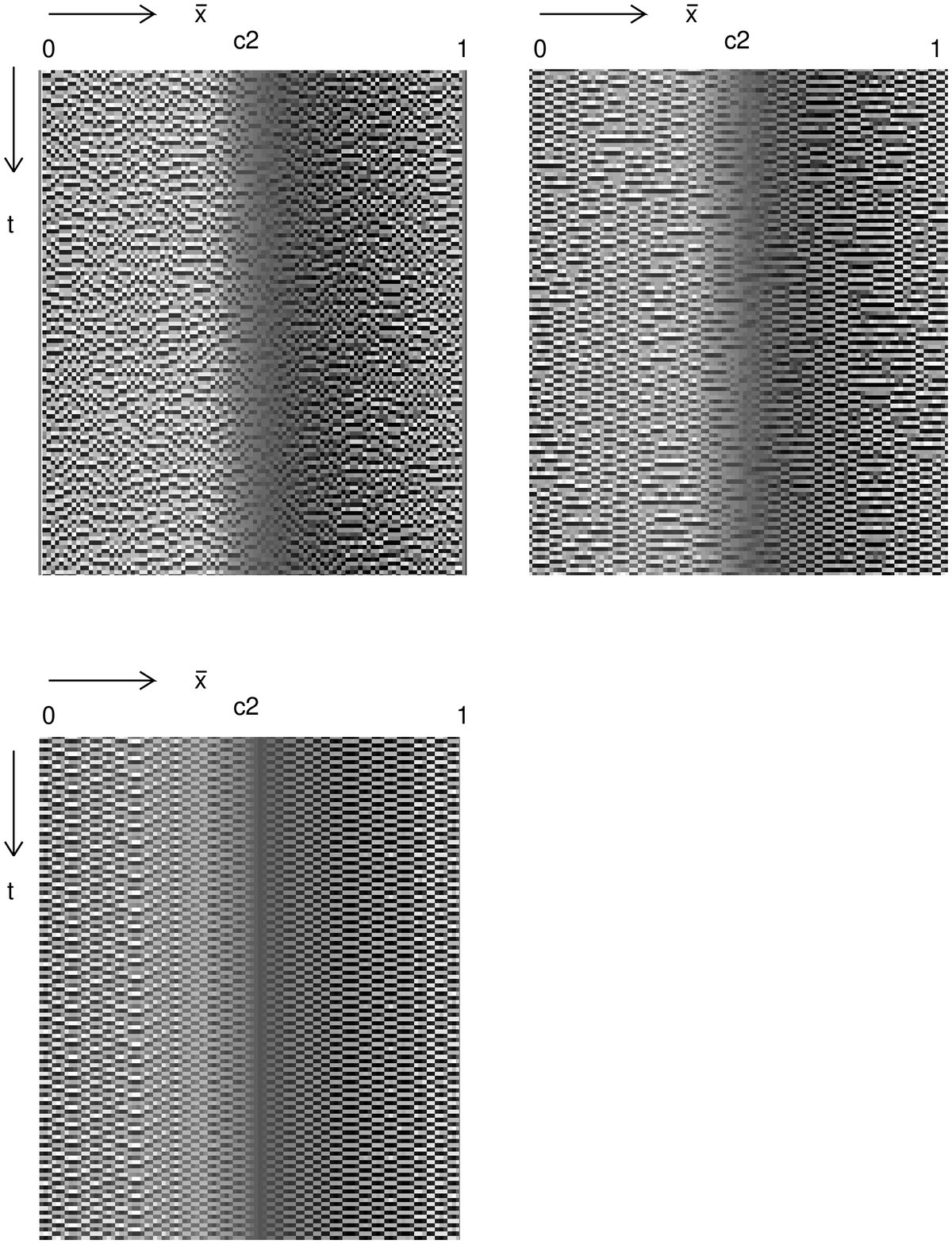,scale=1.0,,
  height =20cm,width =16cm,angle =00}}
\put(-4,-15.){\makebox[8cm]{\centerline{\Large{fig.9}}}}
\put(1, 2.){\makebox[8cm]{\centerline{\Large{a)}}}}
\put(9, 2.){\makebox[8cm]{\centerline{\Large{b)}}}}
\put(1,-9.){\makebox[8cm]{\centerline{\Large{c)}}}}
\end{picture}

\end{document}